\documentstyle[12pt]{article}
\input epsf.tex

\begin{document}
\rightline{McGill/95--12}
\rightline{hep-ph/9503311}
\rightline{(March, 1995)}
\bigskip
\begin{center}
{\Large\bf Diagrammatic Analysis of QCD Gauge Transformations and Gauge
Cancellations}\\
\bigskip\bigskip\bigskip
{Y.J. Feng$^*$ and C. S. Lam$^{\dag}$}\\
\bigskip
{\it Department of Physics, McGill University,\\
3600 University St., Montreal, P.Q., Canada H3A 2T8}
\end{center}
\bigskip\bigskip
\begin{abstract}
Diagrammatic techniques are invented to implement QCD gauge transformations.
These techniques can be used to discover how gauge-dependent terms
are cancelled among diagrams to yield gauge-invariant results in the sum.
In this way a multiloop pinching technique can be developed to change
ordinary vertices into background-gauge vertices.
The techniques can also be used to design new gauges to simplify calculations
by reducing the number of gauge-dependent terms present in the
intermediate steps. Two examples are discussed to illustrate this aspect of the
applications.
\end{abstract}

\section{Introduction}
A typical computation in the Standard Model generates many gauge-dependent
terms that get cancelled at the end. The labour of calculation
can be thus considerably reduced if suitable gauges are chosen to minimize the
presence of these terms.
As far as propagators goes it is usually simplest to use the
Feynman gauge. As to vertices and external
gluon wave functions, more unconventional gauges can often lead
to greater simplifications.

The spinor helicity technique \cite{sht1} is a case in point. External gluon
wave functions
are chosen in lightcone gauges defined by  lightlike reference momenta $k$,
which may be different for different gluons.
A judicious choice of $k$'s can reduce the number of terms present, and
sometimes even renders whole diagrams zero.
This technique was originally designed \cite{sht2,sht3}
for tree-level calculations but with superstring \cite{sust} and
first-quantized \cite{fq}
techniques it can be
extended at least to one-loop diagrams,
and with Schwinger representation it can be extended to multiloops \cite{sr}.

Further simplifications might be obtained by choosing gauges that affect
the vertices. The best known example of this kind is
probably the background
gauge (BG) \cite{bfm}, in which a  triple-gluon (3g) vertex attached to
an external gluon contains
four terms, two less than the six terms present in a normal 3g
vertex. This gauge is particularly convenient for
one-loop $n$-gluon 1PI amplitudes, where each 3g vertex present is such
a BG 3g vertex. In addition to this reduction in number,
of the four terms in such a vertex,
only one involves the internal line and needs to be integrated over.
For other diagrams a different gauge may be more convenient. For example,
in some sense the Gevais-Neveu gauge \cite{gn} is the simplest one for
$n$-gluon
diagrams in the tree approximation. It is quite remarkable that in tree
and one-loop order, the superstring formalism automatically chooses in
some sense these best gauges to compute \cite{sust}.

{}From these examples it is clear that the most suitable gauge to use
depends on the process and the details of the Feynman diagrams. In order to
devise new gauges suitable for a new set of Feynman diagrams, a systematic
study of the mechanism for the cancellation of gauge-dependent terms is needed.
Since Feynman
diagrams are much simpler and more intuitive to visualize than the
corresponding analytic expressions, it would be best for the same reason
if such gauge cancellations can be cast in diagrammatic languages.
This is what we intend to develop in the first part of this article.

How this task is accomplished is well known in QED but not in QCD.
In QCD, a one-loop {\it pinching technique} \cite{pt1,pt2}
is known to simplify calculations
by converting ordinary vertices to BG vertices, though
to our knowledge a general systematic study  for multiloop
is not available.
There are two reasons why cancellation of
gauge-dependent terms and gauge transformations
are considerably more complicated in QCD than in QED.
The first is the complication of color. Fortunately this can be
sidestepped by a color
decomposition and the use of color-oriented diagrams, as will be discussed
in \S  IIA. The second complication is more substantial, and it relates to
the problem of source diffusion. Unlike QED where the charge always resides
on the electron lines, the presence of triple and four-gluon vertices
spread the color globally throughout the Feynman diagram.
In other words, it is the covariant divergence of the color current that
is now zero, and not the usual divergence.
This global nature
means that the local cancellation of gauge-dependence in QED is
no longer sufficient for QCD.
This additional complication of source diffusion fortunately can be handled
by introducing `propagating diagrams', as will be discussed in \S IIB.

One may also view the discussion below in another light.
Imagine starting out from a classical Yang-Mills theory
or a tree-order scattering amplitude. Ghost vertices are absent there.
Suppose now loop diagrams are built up via generalized unitarity
by gluing together tree diagrams. Ghost loops would still be absent after the
gluing so
why then do we need them? The reason must be that without
them the loop amplitudes will no longer be gauge invariant, but how
do we see that? Part of what is being discussed below (Figs.~13--16) can be
thought of as a way of seeing that. Alternatively one may think of what is
being done below as just another way of deriving the BRST transformation, but
this time microscopically, done vertex by vertex and
diagram by diagram.

The basic technique for separating the color and for the creating of the
`propagating diagrams' will be discussed in the next section, throughout which
the gluon propagator is taken to be in the Feynman gauge. This technique
is then used to prove {\it diagrammatically}   known field-theoretic
results. In particular
covariant-gauges will be taken up in \S III, and
{\it multiloop} pinching technique will be discussed in \S IV, where we shall
show
how to
create background gauge vertices from ordinary
vertices by the {\it incomplete cancellation} of divergent parts.
In \S V, we shall discuss two examples for using these techniques
 to design new gauges. These are just examples and there is no
claim that the new gauges are the simplest possible. Nevertheless,
it does illustrate that {\it simpler} gauges can be designed, and hopefully by
working harder one can one day design the `simplest' gauge to be used
for a given set of diagrams. This latter problem is being studied.

\section{Diagrammatic Analysis in the Feynman Gauge}

\subsection{Color-oriented vertices}

If a QCD Feynman amplitude ${\cal T}$ is given a color decomposition into
a set of independent color tensors ${\cal C}_i$, ${\cal T}=\sum_i
{\cal C}_ia_i$, then each of the color-independent subamplitudes $a_i$ is
known to be gauge
invariant [1], though how this is achieved diagrammatically,
so that individual gauge-dependent diagrams add up to give
gauge-independent results, is much less known. In order to study this we must
first learn how to compute the subamplitudes $a_i$ diagrammatically.

The relevant diagrams to compute $a_i$ from are the {\it color-oriented
diagrams}. They differ from the ordinary Feynman diagrams in having fixed
{\it cyclic ordering} of the gluon and the ghost lines at all the vertices.
Vertices with such ordering imposed are the {\it color-oriented vertices};
Feynman diagrams made up of color-oriented vertices are the color-oriented
diagrams\cite{sr}. A single Feynman diagram gives rise to many color-oriented
diagrams, differing from one another in the cyclic ordering of the gluon and
ghost lines at the vertices. Color-oriented vertices arise from ordinary
triple-gluon and ghost vertices by a decomposition of their color factor,
$f^{abc}=-i[{\rm Tr} (T^aT^bT^c)-{\rm Tr}(T^aT^cT^b)].$ This produces
two trace terms with fixed cyclic ordering of the fundamental color generators
$T^a$, corresponding to the two color-oriented vertices. Similarly, the
decomposition of the color factor
$f^{abcd}=(-i)^2[{\rm Tr}(T^aT^bT^cT^d)-{\rm Tr}%
(T^bT^aT^cT^d)-{\rm Tr}(T^aT^bT^dT^c)+{\rm Tr}(T^bT^aT^dT^c)]$ generates
color-oriented vertices for an ordinary four-gluon vertex.
The color factor at a fermion vertex is given by $T^a$. It cannot be further
decomposed so there is only one color-oriented vertex per ordinary fermion
vertex. In that case it does not matter where the gluon line is drawn wrt
the quark line.

In an $U(N)$ theory, each color-oriented diagram gives rise to only
one color tensor ${\cal C}_i$, determined entirely by the cyclic ordering
of the external lines. Thus each $a_i$ is given by the sum of color-oriented
diagrams with a fixed external-line ordering.
For $SU(N)$ theories, a color-oriented diagram may
contain more than one color tensor ${\cal C}_i$, but in what follows we
shall only consider those $a_i$ given by the sum of all
color-oriented diagrams with a fixed ordering of the external lines.

\begin{figure}
\vskip -0 cm
\centerline{\epsfxsize 4.7 truein \epsfbox {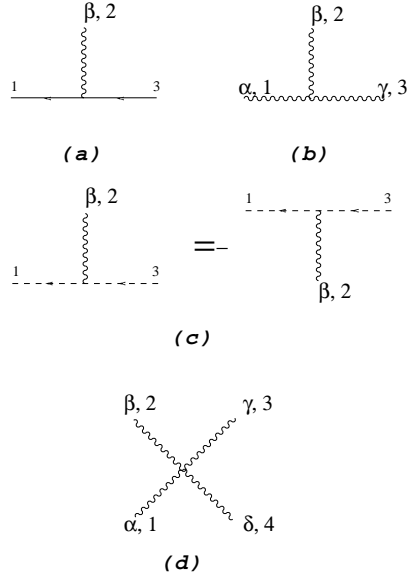}}
\nobreak
\vskip -8.5 cm\nobreak
\vskip .1 cm
\caption{Color-oriented vertices for QCD. }
\end{figure}

With color thus factored out, vertex and propagator factors depend on spin
and momentum but no longer color. The propagators are the usual ones, being
$-1/(\gamma\cdot p-m), -1/p^2$, and $g^{\alpha\beta}/p^2$ respectively for
fermions, ghosts, and gluons.
Up to a sign the color-oriented vertices coincide with the
ordinary vertices without their color factors. The color-oriented vertices are
displayed in Fig.~1; their analytical expressions are given below.

\begin{eqnarray}
F_{\beta}=&&g\  \gamma_{\beta} \ ,\\
T_{\alpha\beta\gamma}(p_1,p_2,p_3)=&&g\
[g_{\alpha\beta} (p_1-p_2)_{\gamma}+
g_{\beta\gamma}(p_2-p_3)_{\alpha}+g_{\gamma\alpha}(p_3-p_1)_{\beta}]
\ , \\
G_{\beta}(p_1)=&&g\  (p_1)_{\beta} \ ,\\
Q_{\alpha\beta\gamma\delta}=&&g^2\  \left[ 2 g_{\alpha\gamma}
g_{\beta\delta}-g_{\alpha\beta}g_{\gamma\delta}-g_{\alpha\delta}
g_{\beta\gamma}\right] \ .
\end{eqnarray}
All momenta in these formulas are outgoing, except those along the quark lines
where they follow the directions of the fermionic arrows.
The lines in Fig.~1c may have two possible ordering, as shown. We shall
occasionally refer to the one with a positive sign, as given by eq.~(1c),
to have the {\it right orientation}, and the one with a minus sign to be of
the {\it wrong orientation}.

\subsection{Divergence relations}

A gauge transformation changes the longitudinal polarization of a gluon.
The corresponding change to a subamplitude is obtained by computing the
divergence of a color-oriented diagram, which in turn is obtained by
computing the divergence of the color-oriented vertices. These divergences are
obtained from eqs.~(2.1) to (2.4) to be
\begin{eqnarray}
(p_2)^{\beta} F_{\beta}&&=g\
\left[-(\gamma\cdot p_1-m)+(\gamma\cdot p_3-m)\right] \ , \\
(p_2)^{\beta} T_{\alpha\beta\gamma}(p_1,p_2,p_3)&&=
g\ g_{\alpha\gamma} p_1^2-g\ g_{\alpha\gamma}p_3^2\
-(p_1)_{\alpha}\  G_{\gamma}(p_1)
+(p_3)_{\gamma}\  G_{\alpha}(p_3) \ ,
\end{eqnarray}
and
\begin{eqnarray}
&&(p_2)^{\beta} G_{\beta}(p_1)+(p_3)^{\gamma}
G_{\gamma}(p_1)+g\ p_1^2 =0 \ , \\
&&(p_2)^{\beta} Q_{\alpha\beta\gamma\delta}-g\ T_{\alpha\gamma\delta}
(p_1+p_2,p_3,p_4)
+g\ T_{\alpha\gamma\delta}(p_1,p_2+p_3,p_4)=0  \ .
\end{eqnarray}

\begin{figure}
\vskip -0 cm
\centerline{\epsfxsize 4.7 truein \epsfbox {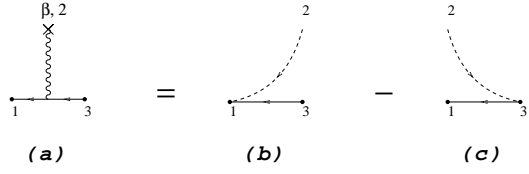}}
\nobreak
\vskip -13 cm\nobreak
\vskip .1 cm
\caption{Divergence relation for the gluon-quark vertex. }
\end{figure}

\begin{figure}
\vskip -0 cm
\centerline{\epsfxsize 4.7 truein \epsfbox {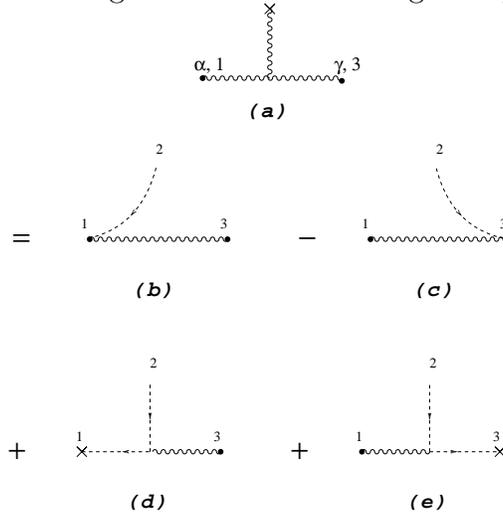}}
\nobreak
\vskip -10 cm\nobreak
\vskip .1 cm
\caption{Divergence relation for the triple gluon vertex. }
\end{figure}

\begin{figure}
\vskip -0 cm
\centerline{\epsfxsize 4.7 truein \epsfbox {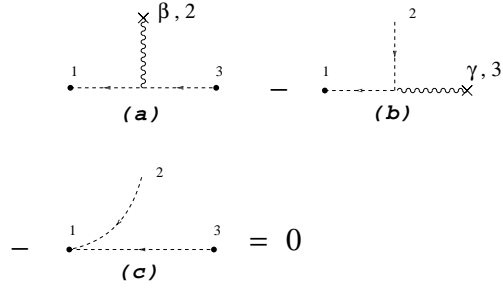}}
\nobreak
\vskip -13 cm\nobreak
\vskip .1 cm
\caption{Divergence relation for the ghost vertex. }
\end{figure}

\begin{figure}
\vskip -0 cm
\centerline{\epsfxsize 4.7 truein \epsfbox {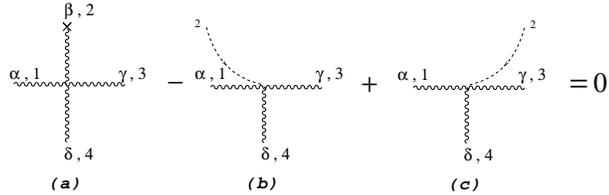}}
\nobreak
\vskip -13.5 cm\nobreak
\vskip .1 cm
\caption{Divergence relation for the four-gluon vertex. }
\end{figure}

The resulting terms have been arranged to be proportional either to a
propagator
or a vertex so that these relations can be expressed diagrammatically, as
in Figs.~2 to 5. {\it A cross in a gluon line represents the divergence},
{\it i.e.,} a factor $p_\alpha$ for the gluon with outgoing momentum
$p$ and Lorentz index $\alpha$. A cross at the end of a (dotted) ghost
line is meant to be a cross on the gluon line it is connected to. Namely,
if $p$ is the outgoing momentum of the ghost line so that $-p$ is the
outgoing momentum and $\alpha$ the Lorentz index of the gluon line it
is connected to, then a cross at the end of the ghost line indicates a
factor $-p_\alpha$.
The propagators in Figs.~2b, 2c, 3b, 3c, 4c, 5b, and 5c have been cancelled
out to obtain these {\it sliding diagrams}, so called because they can be
obtained by converting the original gluon
line into a sliding external ghost (dotted)
line. Diagrams 3d and 3e serve to
propagate the cross along the original gluon lines and will thus be called the
{\it propagating diagrams}. A propagating cross always drags a ghost line
behind it replacing the original gluon line.

QED requires only Fig.~2.
What makes QCD complicated
is the presence of many more vertices and divergence relations,
and the existence of these propagating diagrams relating to the
source diffusion problem discussed in the Introduction.

Possibly with the exception of Fig.~4, these divergence relations have
a regular structure which we shall call the {\it canonical structure}.
On the rhs of these divergence relations there are always two sliding
diagrams with opposite signs, arranged so that the signs obtained from a
triple-gluon
vertex are opposite to the signs obtained from a four-gluon vertex.
There are no propagating diagrams for the four-gluon vertex,
but both of the propagating diagrams emerging from the triple-gluon divergence
relation carry a $+$ sign. It turns out that it is this regular canonical
structure that guarantees the gauge invariance of the sum of diagrams.

In the sliding diagrams, the external ghost line serves to
inject a coupling-constant factor $g$ as well as
the momentum of the original gluon ($p_2$) into the line it is tangential
to, but otherwise it is inert. Thus if
the ghost slides into a momentum-independent
vertex, or is tangential to a line whose momentum the vertex does not depend
on, then the new vertex with the extra ghost line is equal to the old vertex
without it, as shown in Figs.~6, 7, and 8. However, triple-gluon vertices are
momentum-dependent so Figs.~5b and 5c are
not identical, though their difference is simply given by 5a.

\begin{figure}
\vskip -0 cm
\centerline{\epsfxsize 4.7 truein \epsfbox {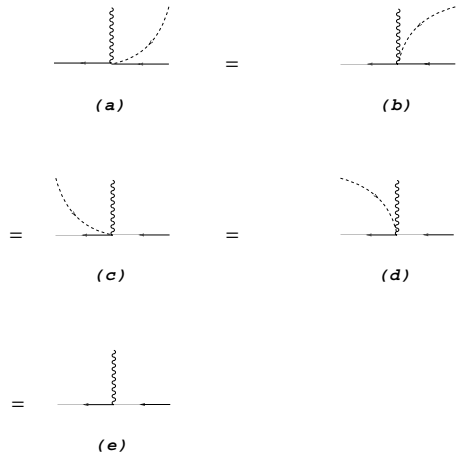}}
\nobreak
\vskip -10 cm\nobreak
\vskip .1 cm
\caption{Relations between sliding diagrams. }
\end{figure}

\begin{figure}
\vskip -0 cm
\centerline{\epsfxsize 4.7 truein \epsfbox {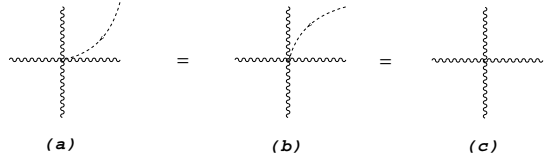}}
\nobreak
\vskip -13.5 cm\nobreak
\vskip .1 cm
\caption{Diagrams with a ghost line sliding into a four gluon vertex. }
\end{figure}

\begin{figure}
\vskip -0 cm
\centerline{\epsfxsize 4.7 truein \epsfbox {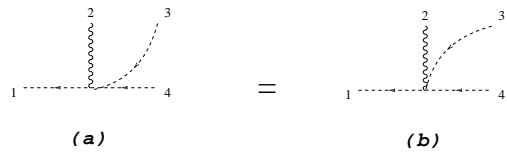}}
\nobreak
\vskip -14 cm\nobreak
\vskip .1 cm
\caption{Ghost vertex with an extra ghost sliding in. }
\end{figure}

The signs appearing in Figs.~3d and 3e require an explanation. First, the
crosses in these diagrams are respectively $-(p_1)_\alpha$ and $-(p_3)_\gamma
$. Secondly, the original gluon {\it propagators} for line 1 in Fig.~3d,
and line 3 in 3e, are now replaced by ghost {\it propagators}, resulting in an
extra minus sign  for both. Thirdly, an extra minus sign is attached to
the ghost vertex
in Fig.~3d because of its wrong orientation.
Putting these three things together, the signs in front of 3d and
3e are both $+$, a fact which becomes very important later on for ghost
cancellation.

Eqs.~(2.5) and (2.6) (Figs.~2 and 3) can be used
repeatedly to  compute the divergence of a
color-oriented diagram. This iteration terminates when
the cross either rests on (i) an external
line, (ii) a four-gluon vertex like Fig.~5a, or (iii) a ghost vertex like
Fig.~4a, or that (iv) there is no longer any cross in the diagram.
If (iv) happens, it is possible for the sliding ghost line to end up (iv a) at
another external line, (iv b) at a fermion vertex like Fig.~6, (iv c) at a
four-gluon vertex like Fig.~7,  (iv d) at a ghost vertex like Fig.~8, or
(iv e) at a three-gluon vertex like Fig.~5b
or 5c.

\subsection{Notations and conventions}

By a `diagram' we always mean a color-oriented diagram from now on.
By the `sum of all diagrams' we always mean the sum of all color-oriented
diagrams with the same ordering of the external lines as the original
diagram.

There are more diagrams in this paper than equations.
In order not to confuse diagram numbers with equation numbers, we will adhere
to the convention that a number not prefixed by Fig. or eq. is taken to
mean a {\it diagram number}.

A diagram is said to be {\it on-shell} if all its external lines are on-shell.
It is said to be {\it on-shell/crossed} if all its external lines are
either on-shell, or are off-shell gluon lines carrying a cross.
Most of the following results apply either to on-shell diagrams, or
on-shell/crossed diagrams.

\subsection{Gauge transformation of color-oriented diagrams}

We consider now how a color-oriented diagram changes  when the
wave function of an external gluon line  undergoes a gauge variation. The
 {\it change}  is proportional to the divergence, represented graphically
by a cross at the end of the external gluon line. This produces many diagrams
by using 2 and 3,
each satisfying one of (i) to (iv) at the end of \S IIB.

We shall  now show that the {\it net change} from
the sum of all diagrams vanishes if the diagrams are on-shell. This occurs
because each resulting diagram either vanishes by itself, or they combine
to cancel each other in pairs or in threesome.
In the process
of the proof we will also discover what remains if
some of the external lines are taken off-shell. These will turn out to be
precisely those diagrams required by the BRST transformations.

These conclusions are certainly of no surprise. What we gain by carrying
out these
analyses is the knowledge how this is realized diagrammatically.

It is simplest to consider first diagrams satisfying condition (iv). When
(iv a) is satisfied, the external ghost
ends on another external line. The pair of ghost/external lines
no longer possesses a particle pole to overcome the Klein-Gordon zero in front
of the LSZ reduction formula, so such diagrams do not contribute to on-shell
scattering amplitudes. If the external line is off-shell
this diagram survives, as is demanded by the BRST transformation.

\begin{figure}
\vskip -0 cm
\centerline{\epsfxsize 4.7 truein \epsfbox {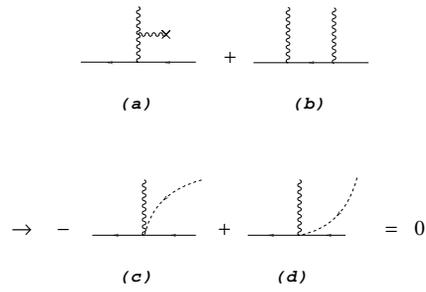}}
\nobreak
\vskip -12 cm\nobreak
\vskip .1 cm
\caption{Local gauge cancellation near a quark vertex.}
\end{figure}

\begin{figure}
\vskip -0 cm
\centerline{\epsfxsize 4.7 truein \epsfbox {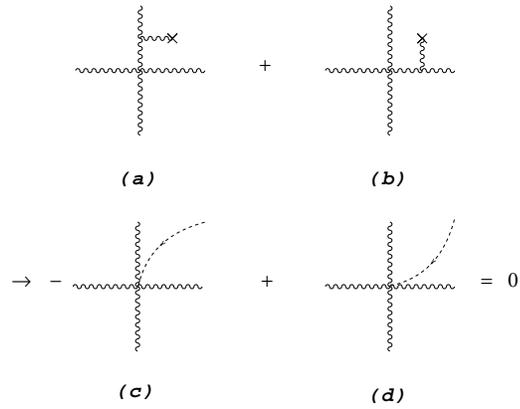}}
\nobreak
\vskip -11 cm\nobreak
\vskip .1 cm
\caption{Local gauge cancellation near a four-gluon vertex. }
\end{figure}

\begin{figure}
\vskip -0 cm
\centerline{\epsfxsize 4.7 truein \epsfbox {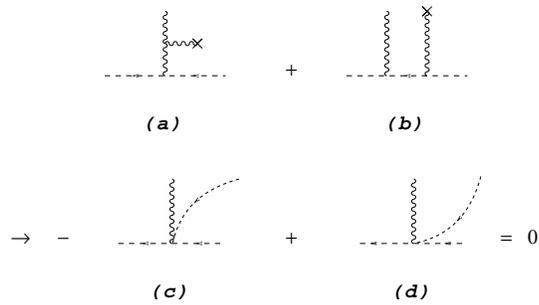}}
\nobreak
\vskip -12 cm\nobreak
\vskip .1 cm
\caption{Local gauge cancellation near a ghost vertex.}
\end{figure}

Next, suppose (iv b) happens. Then using Fig.~6 a pairwise cancellation takes
place as shown in Fig.~9, where diagrams 9a and 9b to the left of the
arrow are the diagrams from which 9c and 9d come from one step back in the
transformation. Note that Fig.~9, and similarly for all the diagrams
considered below, is meant to be a part of a much larger color-oriented
diagram, and not necessarily a whole diagram by itself.

Similar pairwise cancellation occurs for (iv c) and (iv d) as shown in
10 and 11. Using 5, case (iv e) combines with case (ii) to give a
threesome cancellation as shown in Fig.~12.

Strictly speaking, there is one more case under (iv d) which we have not yet
considered, namely, when the external ghost line slides into a quadrant
bounded by the {\it outgoing} ghost line, as shown in 15a and 15b.
This case will be considered at the end. There are two reasons for the
asymmetry between this case and 11, both arising from the asymmetry of
the ghost vertex, which depends on the momentum of the outgoing ghost but
not the incoming ghost, nor the gluon. As a result, the sliding term in
4 slides only to the left but not to the right. Moreover, Fig.~8 is
true, but there is no corresponding relation when the external sliding ghost
lies between the gluon and the {\it outgoing} ghost.

We must now deal with cases where the external ghost line ends at a cross.
Case (ii) has already been dealt with but we still have to consider cases
(i) and (iii). Case (i) is simple. If that external line the cross ends on
is on-shell, then the amplitude vanishes because external gluon wave
functions are divergenceless. Otherwise this diagram remains as required by
the BRST transformations.

\begin{figure}
\vskip -0 cm
\centerline{\epsfxsize 4.7 truein \epsfbox {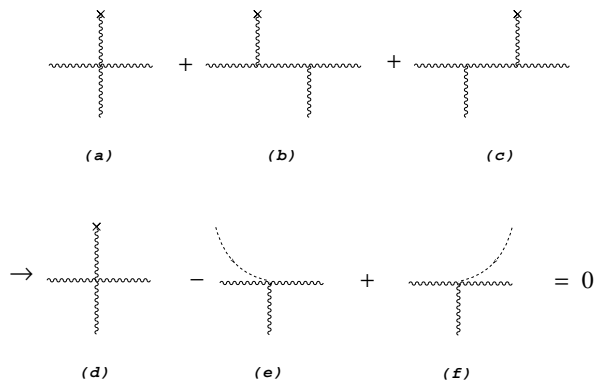}}
\nobreak
\vskip -11 cm\nobreak
\vskip .1 cm
\caption{Local gauge cancellation near a three-gluon vertex.}
\end{figure}

\begin{figure}
\vskip -0 cm
\centerline{\epsfxsize 4.7 truein \epsfbox {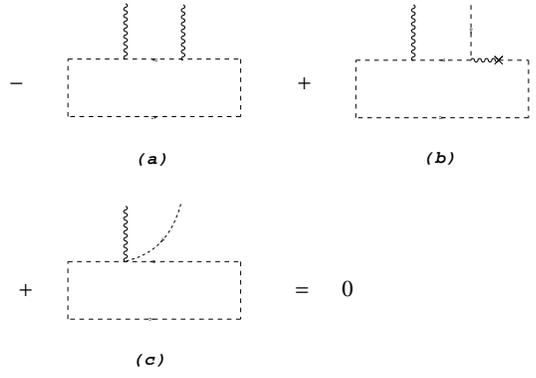}}
\nobreak
\vskip -11 cm\nobreak
\vskip .1 cm
\caption{Cancellation involving ghost loop. Same as Fig.~4 with ghost
loop explicitly drawn in.}
\end{figure}

\begin{figure}
\vskip -0 cm
\centerline{\epsfxsize 4.7 truein \epsfbox {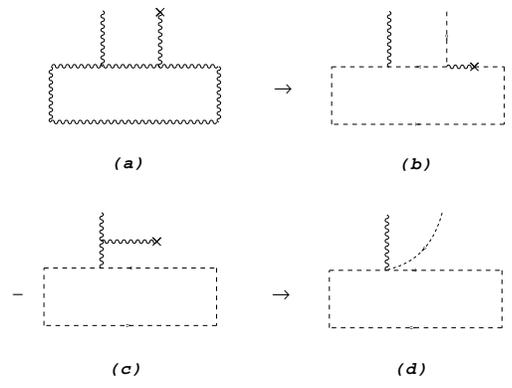}}
\nobreak
\vskip -11 cm\nobreak
\vskip .1 cm
\caption{How 13b and 13c are produced from color-oriented diagrams.}

\end{figure}

\begin{figure}
\vskip -0 cm
\centerline{\epsfxsize 4.7 truein \epsfbox {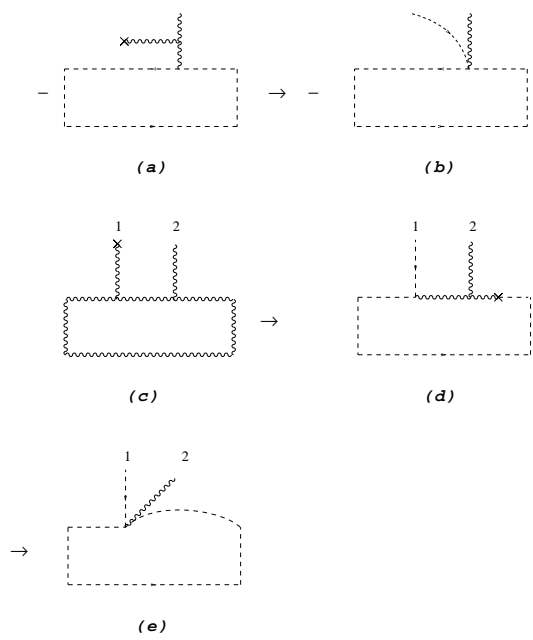}}
\nobreak
\vskip -8.5 cm\nobreak
\vskip .1 cm
\caption{How 16a and 16b are produced from color-oriented diagrams.}
\end{figure}

\begin{figure}
\vskip -.5 cm
\centerline{\epsfxsize 4.7 truein \epsfbox {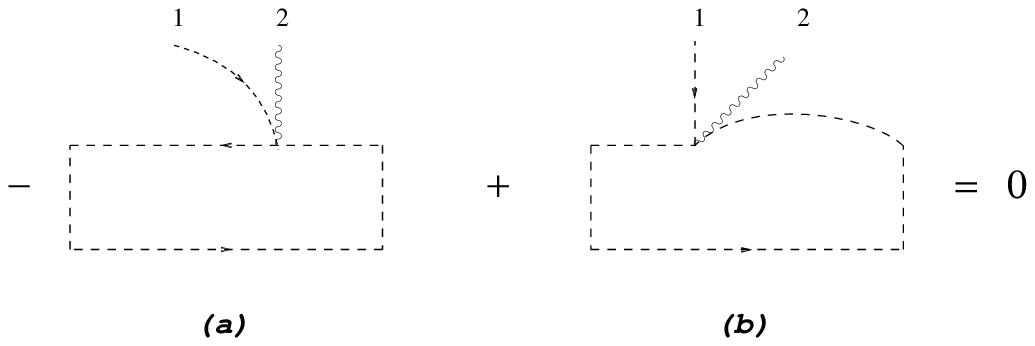}}
\nobreak
\vskip -13 cm\nobreak
\vskip .1 cm
\caption{Pairwise cancellation for the other situation of case (iv d).}
\end{figure}

The identity in Fig.~4 is required to deal with case (iii).
It is important in this connection to note that
the ghost line appearing in 4a must be an internal ghost line,
forming a closed loop as in 13a. This means that the identity in 4 will always
look like the identity in 13, so
the only question left is where 13b and 13c will
come from. With the presence of the ghost loop in 13a, there must be a diagram
where the ghost loop is replaced by a gluon loop, as in 14a. Using 3d,
3e repeatedly on 14a, we shall end up with a diagram that looks like
13b=14b. The minus sign in front of 13a comes from the ghost loop factor
and there is no corresponding minus sign for the gluon loop in 14a.
In addition to 13a, there is a
diagram like it but the gluon with the cross is linked up directly with the
other gluon, as in 14c. The $-$ sign in front is again due to the ghost
loop. Using 3c and then the equality in Fig.~8, 14c produces 14d=13c. The
three diagrams in Fig.~13 cancel one another by using Fig.~4. The same
proof will go through if the arrow of the ghost loop in 13a runs clockwise
instead.

There is one final case which we have not yet dealt with, namely, the other
situation of case (iv d) which we skipped before. Consider 15a, where
the explicit $-$ sign in front is again the ghost loop factor. Using 3b,
this is transferred to 15b=16a, with the external ghost between the gluon
and the {\it outgoing} internal ghost line. Its cancellation comes from the
gluon-loop diagram shown in 15c. Using 3d and 3e, 15c is transformed into
15d, which by 3b, becomes 15e=16b. Now 16a+16b=0 because the vertices in
these two diagrams are simply the same vertex drawn a bit differently.

In summary, we have shown that the sum of all on-shell diagrams with a cross
at the end of a gluon line is zero.

\section{Covariant Gauges}

The gluon propagator in a covariant gauge is $(g^{\alpha\beta}+\xi
p^\alpha p^\beta/p^2)/p^2$. In the last section divergence relations
and gauge invariances of the external gluons were shown in the Feynman
gauge $\xi=0$. The effect of $\xi\not=0$ will be discussed in the present
section, using the same techniques developed in the last.

The only additional tool needed for the following analysis is the
double-divergence relation
\begin{eqnarray}
(p_3)^\gamma(p_2)^\beta T_{\alpha\beta\gamma}(p_1,p_2,p_3)
=&&g\ (p_3)_\alpha p_1^2 \nonumber\\
-&&(p_1)_\alpha(p_3)^\gamma G_\gamma(p_1)\ ,
\end{eqnarray}

\begin{figure}
\vskip -0 cm
\centerline{\epsfxsize 4.7 truein \epsfbox {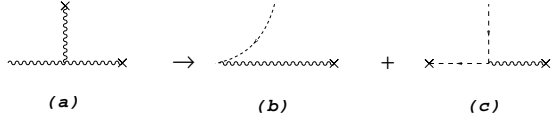}}
\nobreak
\vskip -13 cm\nobreak
\vskip .1 cm
\caption{Double-divergence relation of the triple gluon vertex.}
\end{figure}

obtainable from eq.~(2.6) by taking the divergence wrt $p_3$ on both sides.
Diagrammatically, this can be represented by Fig.~17, which is the same
as Fig.~3 with 3b and 3e removed and with an additional cross added to
the third gluon line. A convenient way to visualize this is to regard Fig.~3
to be true in all cases,
with or without an additional cross on line 3, but to add to that
the additional rule that the ghost line on the rhs must avoid coming into
contact with any additional crosses. This rule effectively eliminates
3b and 3e as required.

This double-divergence relation allows us to generalize the result of
the \S IID into\hfill\break
\bigskip

\noindent{\bf Lemma 1:}\quad {\it The sum of all on-shell/crossed diagrams
with a cross on
one of the external gluon lines is zero, if all gluon propagators are taken in
the Feynman gauge}.\hfill\break

\bigskip
This lemma is more general than what has been proved because
the external lines with a cross on them do not
have to be on-shell, so diagrams of types
(iv a) and (i) with the external ghost resting on them will no
longer vanish. However, such diagrams can never occur because of eq.~(3.1),
which as noted before may be taken to say that the external ghost line must
avoid the additional crosses now situated at the end of the off-shell gluon
lines.

This lemma can be used to show the independence of
the gauge parameter $\xi$.
\hfill\break\bigskip

\noindent{\bf Lemma 2}\quad {\it The sum of on-shell/crossed diagrams
is independent of the gauge choice of every gluon propagator}.
\hfill\break\bigskip

The gauge-dependent part of a gluon propagator is
of the form $\xi p^\alpha p^\beta/p^2/p^2$. Except for the additional factor
$\xi/p^2/p^2$,
the $\xi$-dependent part can be represented diagrammatically
by breaking the internal line  into a pair of external lines, each with
a cross at the end. In this way two diagrams are generated by each gluon
propagator: one being the original diagram with the propagator in the Feynman
gauge, and the other obtained by breaking this propagator into
a pair of external lines with a cross at the end of each.

If a diagram has $m$ propagators, then this procedure breaks it up into
$2^m$ diagrams. When we sum up all possible on-shell/crossed diagrams,
the sum becomes $2^m$ sets of sums, each of which satisfies Lemma 1,
so Lemma 2 is proved.

\section{Pinching Technique and the Background Gauge}

The technique developed in \S II
can also be used to manipulate the divergence part of a
triple-gluon vertex.
Such vertex manipulations had previously been used
to simplify one-loop calculations, and in
that context it is known as the {\it pinching technique} [8]. To one-loop order
it has been shown that pinching technique gives rise to the {\it background
gauge} (BG). In what follows we generalize this approach to {\it multiloops}
to show how to convert the normal vertices into the BG vertices.

\subsection{BG vertices}
The BG color-oriented vertices consist of the original
vertices in Fig.~1 (eqs.~(2.1) to (2.4)),
plus the additional vertices in Fig.~18
given by the following formulas:

\begin{eqnarray}
&&18a=g\ [g_{\gamma\alpha}(p_3-p_1)_\beta-2g_{\alpha\beta}(p_2)_\gamma
+2g_{\beta\gamma}(p_2)_\alpha]\ ,\\
&&18b=g\ (p_1-p_3)_\beta\ ,\\
&&18c=g\ [-g_{\beta\gamma}g_{\alpha\delta}-2g_{\alpha\beta}g_{\gamma\delta}
+2g_{\alpha\gamma}g_{\beta\delta}] ,\\
&&18d=-2g\ g_{\beta\gamma}g_{\alpha\delta}\ ,\\
&&18e=g\ g_{\beta\gamma}\ ,\\
&&18f=g\ g_{\beta\gamma}\ ,\\
&&18g=-2g\ g_{\beta\gamma}\ .
\end{eqnarray}

Gluon lines with an arrow are to be distinguished from gluon lines without
an arrow. In this section we follow the traditional usage to assume all
arrowed lines to be external gluon lines, though the reverse is not true.
Vertices attached to external gluon lines should be taken from 18
whenever possible, but if
they are absent from 18, then they should be taken from 1. For example,
triple-gluon vertices with two external gluons and quark-gluon vertices
with an external gluon are not to be found in 18, so they should be
taken from 1b and 1a respectively.

\begin{figure}
\vskip .8 cm
\centerline{\epsfxsize 4.7 truein \epsfbox {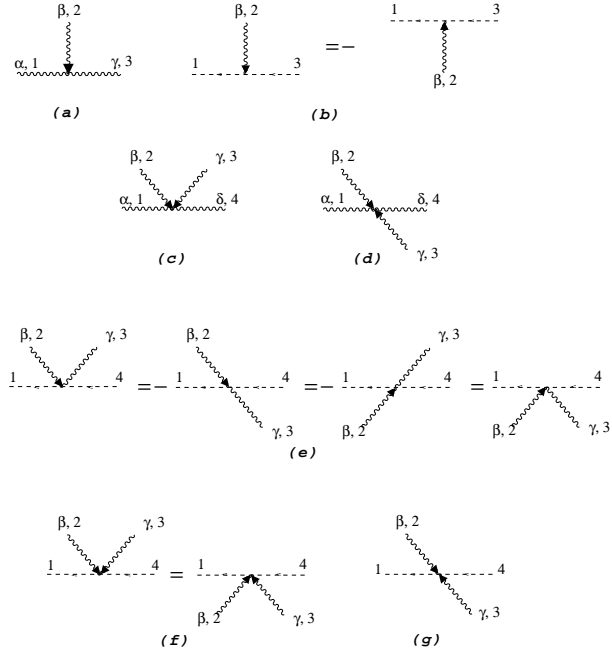}}
\nobreak
\vskip -8.3 cm\nobreak
\vskip .1 cm
\caption{Color-oriented BG vertices with at least one arrowed (external)
line.}
\end{figure}

It is this distinction between external and internal gluons that produces
so many vertices in the BG. In spite of this complication, it is often
still simpler to calculate loop
processes with a large number of external gluon lines in the BG,
because of its reduced dependence on the internal momenta of a triple-gluon
vertex. BG is also the gauge that emerges naturally in superstring calculations
for 1PI diagrams [3].

The equivalence between the ordinary and the BG vertices is stated in the
following lemma.

\subsection{Lemma 3}
{\it The sum of on-shell/crossed diagrams are not affected when
the vertices attached to any number ($n$) of external gluon lines are changed
from the ordinary vertices (Fig.~1) to the BG vertices
(Figs.~1 and 18).}

\subsection{Creation of the new BG vertices}

To prove Lemma 3 we need to know how the new BG vertices in Fig.~18 are
related to the ordinary vertices in Fig.~1.
These relations can be obtained from a relation between
1b and 18a, or equivalently between eqs.~(2.2) and (4.1), shown graphically in
Fig.~19. For this purpose it is necessary to introduce a new vertex with
two gluon and one ghost lines (19b and 19c), joined together by a circle
representing the factor $\pm g\ g_{\mu\nu}$, where $\mu,\nu$
are the Lorentz indices of the two gluon lines. The sign is taken to be
$+$ for 19c and $-$ for 19b. This sign convention is chosen to be
the same as the ordinary ghost vertex 1c if the incoming ghost line
is replaced by the arrowed gluon line. In the terminology introduced
below eq.~(2.4), 19c has the right orientation and
19b has the wrong orientation.
{}From now on we shall refer to the
vertices 19b,c as the {\it funny vertices}.

\begin{figure}
\vskip -0 cm
\centerline{\epsfxsize 4.7 truein \epsfbox {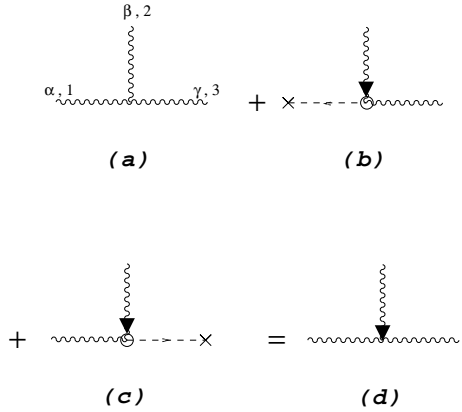}}
\nobreak
\vskip -10.5 cm\nobreak
\vskip .1 cm
\caption{Relation between the triple-gluon vertices 1b and 18a.}
\end{figure}

\begin{figure}
\vskip -0 cm
\centerline{\epsfxsize 4.7 truein \epsfbox {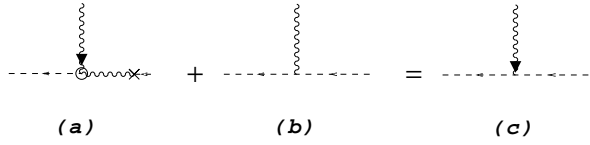}}
\nobreak
\vskip -14 cm\nobreak
\vskip .1 cm
\caption{Generation of the ghost vertex 18b.}
\end{figure}

\begin{figure}
\vskip -6.5 cm
\centerline{\epsfxsize 4.7 truein \epsfbox {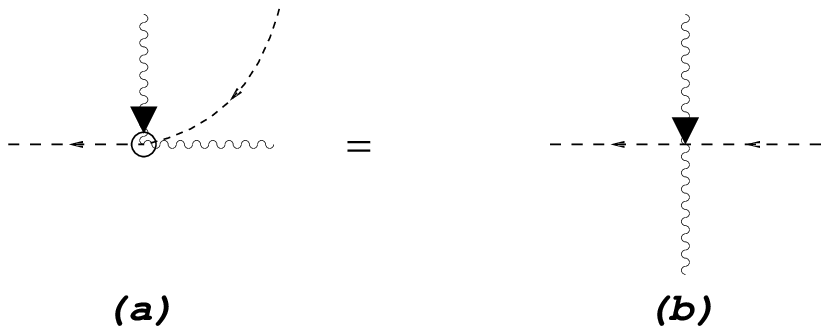}}
\nobreak
\vskip -7.2 cm\nobreak
\vskip .1 cm
\caption{Generating the ghost vertex 18e.}
\end{figure}

\begin{figure}
\vskip -2 cm
\centerline{\epsfxsize 4.7 truein \epsfbox {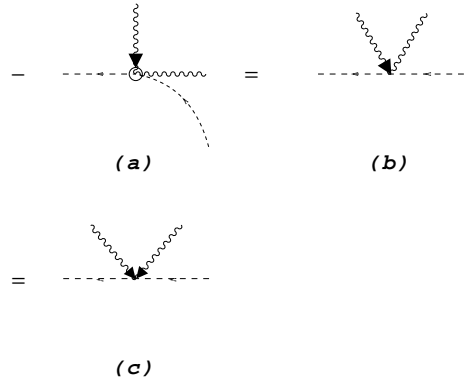}}
\nobreak
\vskip -8.5 cm\nobreak
\vskip .1 cm
\caption{Generating the ghost vertices 18e and 18f.}
\end{figure}

\begin{figure}
\vskip -1.5 cm
\centerline{\epsfxsize 4.7 truein \epsfbox {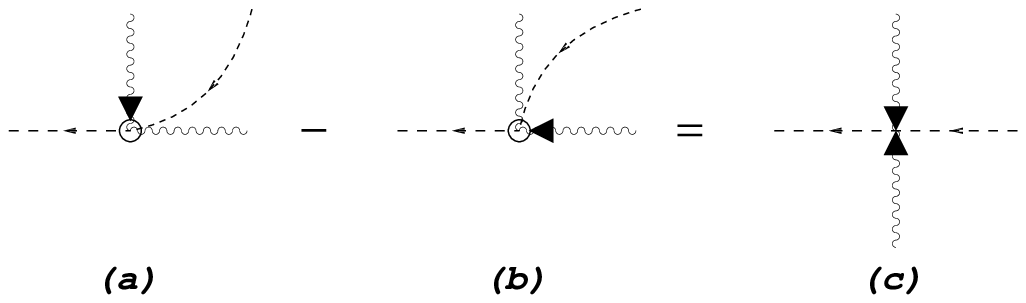}}
\nobreak
\vskip -13 cm\nobreak
\vskip .1 cm
\caption{Generating the ghost vertex 18g.}
\end{figure}

\begin{figure}
\vskip -3 cm
\centerline{\epsfxsize 4.7 truein \epsfbox {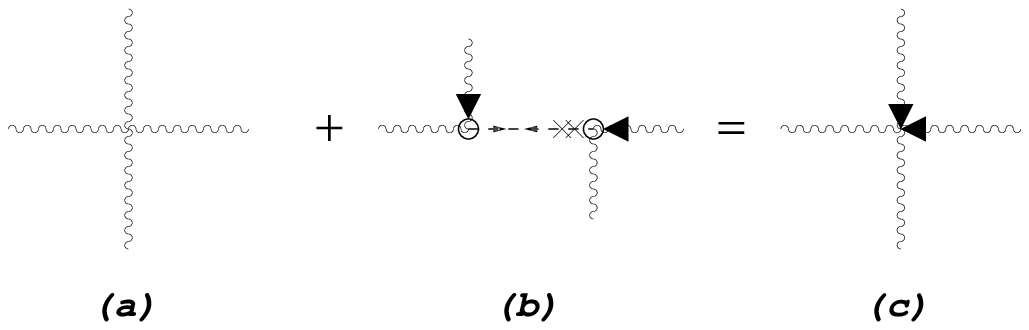}}
\nobreak
\vskip -10.5 cm\nobreak
\vskip .1 cm
\caption{Generating the four-gluon vertex 18c.}
\end{figure}

\begin{figure}
\vskip -1 cm
\centerline{\epsfxsize 4.7 truein \epsfbox {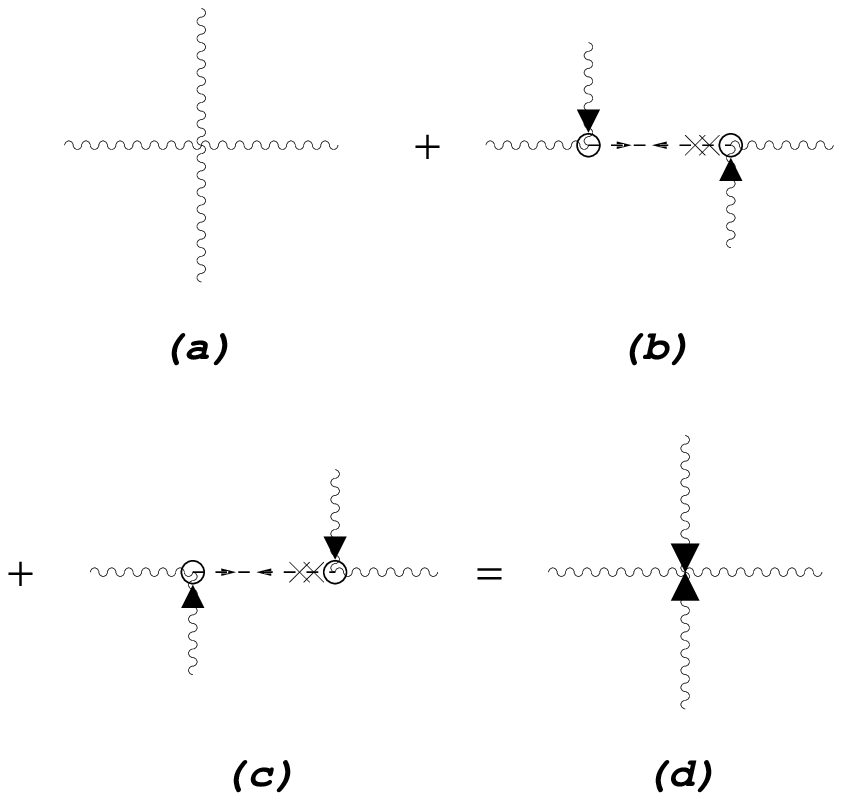}}
\nobreak
\vskip -10.5 cm\nobreak
\vskip .1 cm
\caption{Generating the four-gluon vertex 18d.}
\end{figure}

Fig.~19 is used to convert an ordinary vertex 19a to a BG vertex 19d.
The two extra terms 19b and 19c will be shown to be instrumental in
converting other ordinary vertices into BG vertices. The proof proceeds by
induction on $n$, the number of external lines at which
such conversion of vertices is desired.

Suppose $n=1$ and an ordinary vertex 19a attached
a single external gluon line is converted into the BG vertex 19d.
The effect of the extra term 19b will now be explained, and an analogous
result exists for 19c. If we remove the bulk of the vertex 19b
from a  diagram except
for its cross, Lemma 1 is applicable to the remaining diagram, so
the extra diagrams generated by 19b will add up to zero, except when
the crosses in 19b is returned by propagation to the
same vertex via a gluon loop. In that case 19b gives rise to 20a, in which
the gluon loop linking the two sides of the funny vertex is now converted
to a ghost loop trailing the cross. There exists another diagram where the
gluon
loop is replaced by the ghost loop, in which the ordinary ghost vertex 20b
appears. The combination of 20a and 20b now gives rise to the
BG ghost vertex 20c. An extra minus sign from the ghost loop has been
incorporated in front of 20a to make the combined sign $+$ as shown.
However, analytically 20a contributes a minus sign because of the wrong
orientation of the funny vertex.

The only other way the cross in 19b can return is through a sliding
diagram 3b at the last step, thus producing 21a or 22a. Since the propagating
cross always drags a ghost line behind it, the gluon loop through which
the cross returns has now been changed into a ghost loop and an examination
of eqs.~(4.5) and (4.6) shows that 21a is the
same as 21b=18e and 22a is the same as 22b=18e and 22c=18f.

We have therefore completed showing Lemma 3 when $n=1$, because none
of the other
BG vertices in 18 are present for $n=1$.

We will now proceed to $n=2$ and assume the first vertex to have been
converted already into
BG vertices. We must now examine the effect of converting the second
vertex from ordinary to BG vertices,
again using the relations in 19.
Clearly as in the case $n=1$, there is no
problem is converting the second vertex into BG vertices if it were alone.
But with $n=2$, there is now
the possibility of an interaction between the two vertices to cause a change.

To start with,  we can no longer use Fig.~3 to propagate the cross
beyond the first vertex because the triple-gluon vertex here is already a
BG vertex. We must therefore work out a relation analogous
to 3 but valid for the BG triple-gluon vertex. This is obtained from
eq.~(4.1) and shown in 26, with 26f being $-g\ g_{\alpha\gamma}(p_2)^2$.
It is important to note that this divergence still possesses a regular
structure.
As before, we have the sliding diagrams with opposite signs (26b  and 26c),
the propagating diagrams with plus signs (26d and 26e), but there is now
a new {\it stagnant diagram} involving the funny vertex (26f) with
a single ghost line that goes nowhere. The analytical expression
for this term is $-p_2^2g_{\alpha\gamma}$, with a minus sign on account of
the wrong orientation of the funny vertex.
Note also that the gluon line in 26d has an arrow but not the one in 26e,
`because' that is how it is inherited from 26a.

Although this is getting ahead of ourselves, it would be useful for the
sake of comparison to examine this {\it modified canonical
structure} for other divergence relations obtained from eqs.~(4.1) to
(4.7), and shown in
Figs.~27 to 32. There are unfortunately many divergence relations, but this
cannot be helped because there are
many BG vertices, and because there are many external-line/cross combinations
in taking the divergence of 1d.  Nevertheless, all these relations
possess the sliding diagrams with the canonical signs,
the propagating diagrams with plus signs, and the stagnant diagrams with minus
signs.
External lines are inherited, vertices that do not make sense will not appear,
and in the case of the divergence of a four-gluon vertex, the cross
is not allowed to propagate through an arrowed line. This last rule is
`why' there are no propagating ghosts from the top to the bottom gluon
lines in 27, for example.
The reason `why' there is no propagating ghost from the top to the
left line in
Fig.~29 is because the resulting BG ghost vertex does not exist (the
arrowed line is not adjacent to the outgoing ghost).

\begin{figure}
\vskip -1 cm
\centerline{\epsfxsize 4.7 truein \epsfbox {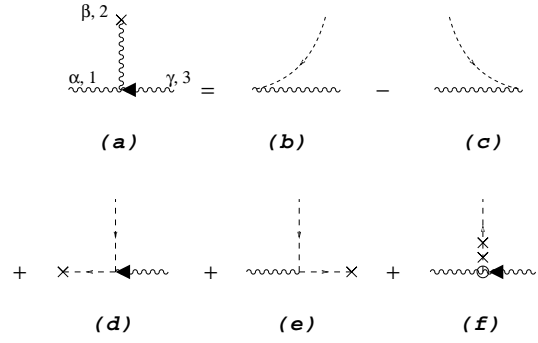}}
\nobreak
\vskip -11.2 cm\nobreak
\vskip .1 cm
\caption{Divergence relation of the BG triple-gluon vertex 18a.}
\end{figure}

\begin{figure}
\vskip 0 cm
\centerline{\epsfxsize 4.7 truein \epsfbox {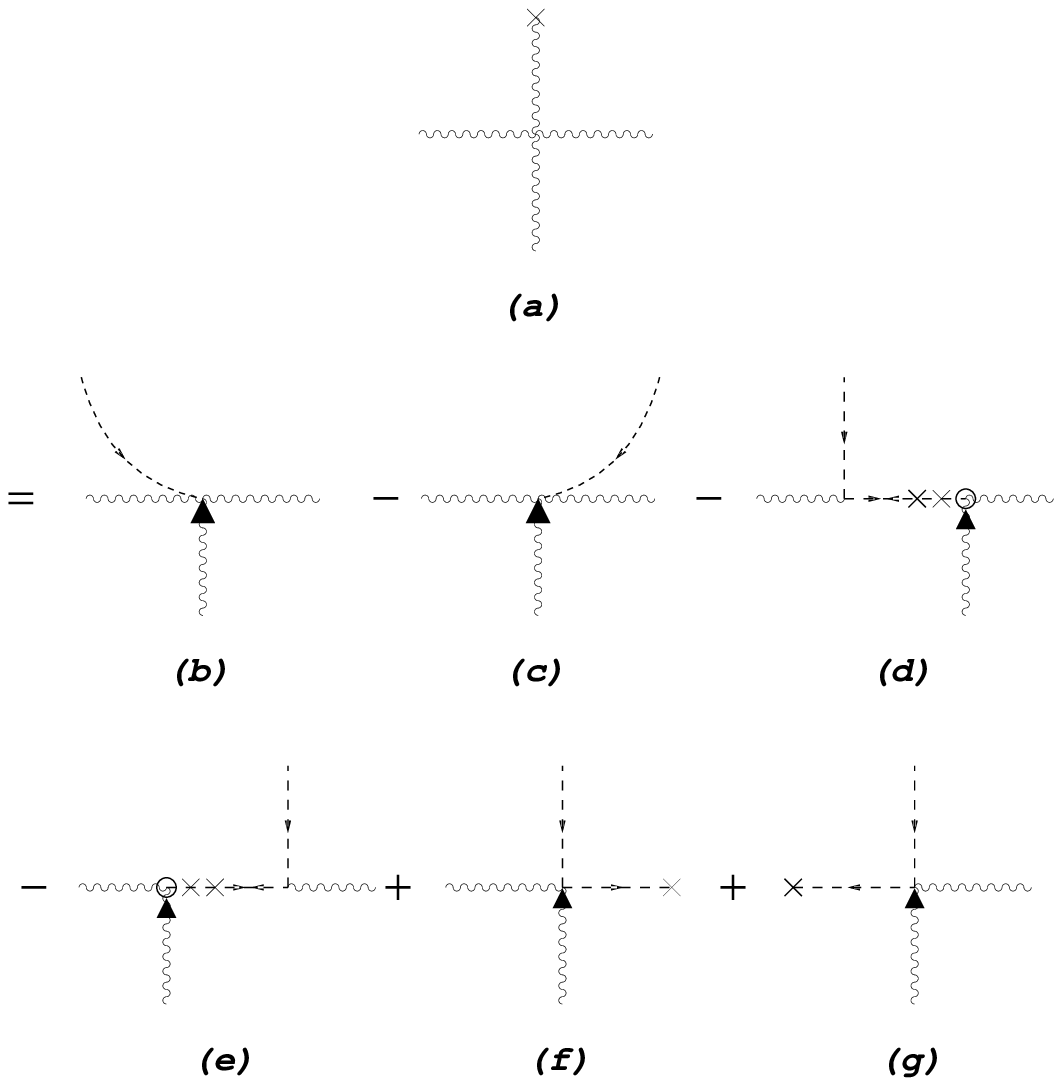}}
\nobreak
\vskip -8 cm\nobreak
\vskip .1 cm
\caption{Divergence relation of the four-gluon vertex 1d expressed in terms of
BG vertices. The cross is opposite to the external line in the four-gluon
vertex.}
\end{figure}

\begin{figure}
\vskip -1 cm
\centerline{\epsfxsize 4.7 truein \epsfbox {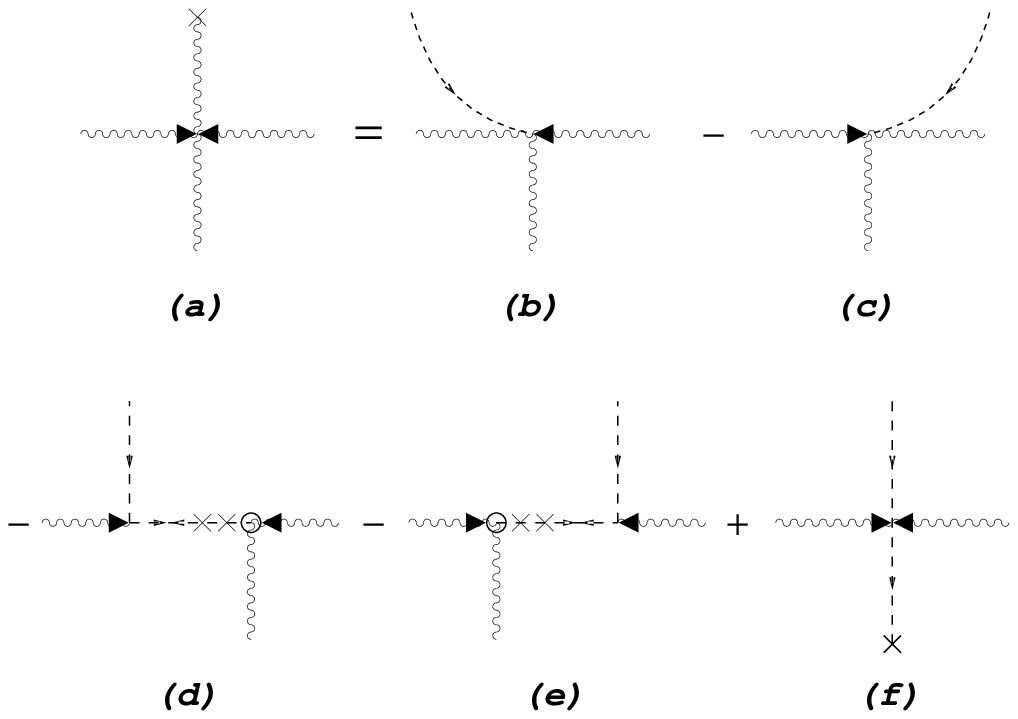}}
\nobreak
\vskip -10.5 cm\nobreak
\vskip .1 cm
\caption{Divergence relation of the four-gluon vertex 18d expressed in terms
of the BG vertices.}
\end{figure}

\begin{figure}
\vskip -3 cm
\centerline{\epsfxsize 4.7 truein \epsfbox {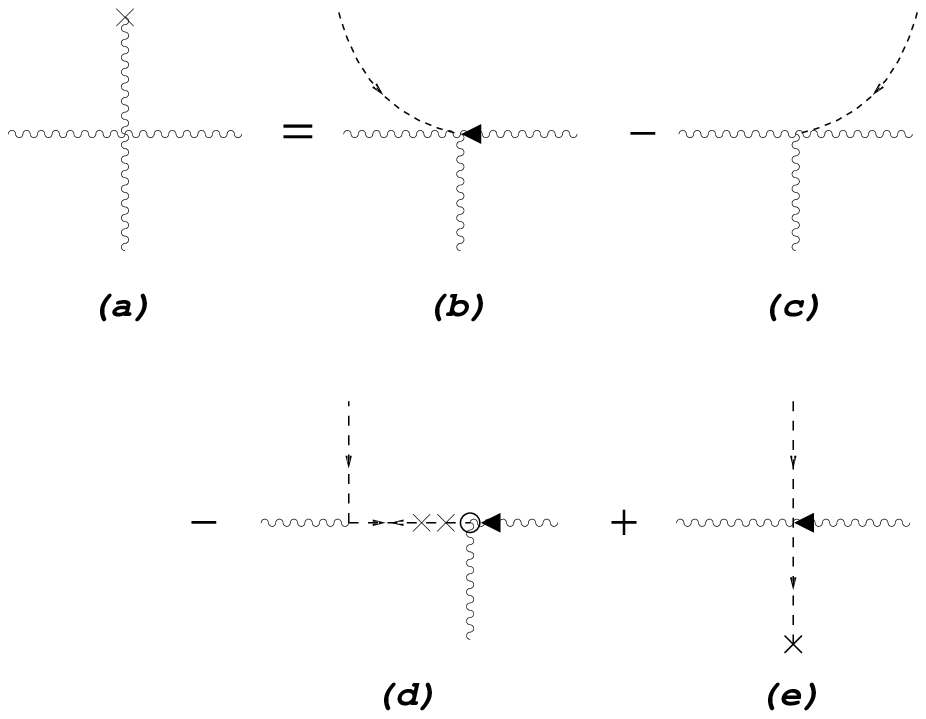}}
\nobreak
\vskip -9 cm\nobreak
\vskip .1 cm
\caption{Divergence relation of the four-gluon vertex 1d expressed in terms of
BG vertices. The cross is adjacent to the external line in the four-gluon
vertex.}
\end{figure}

\begin{figure}
\vskip -0 cm
\centerline{\epsfxsize 4.7 truein \epsfbox {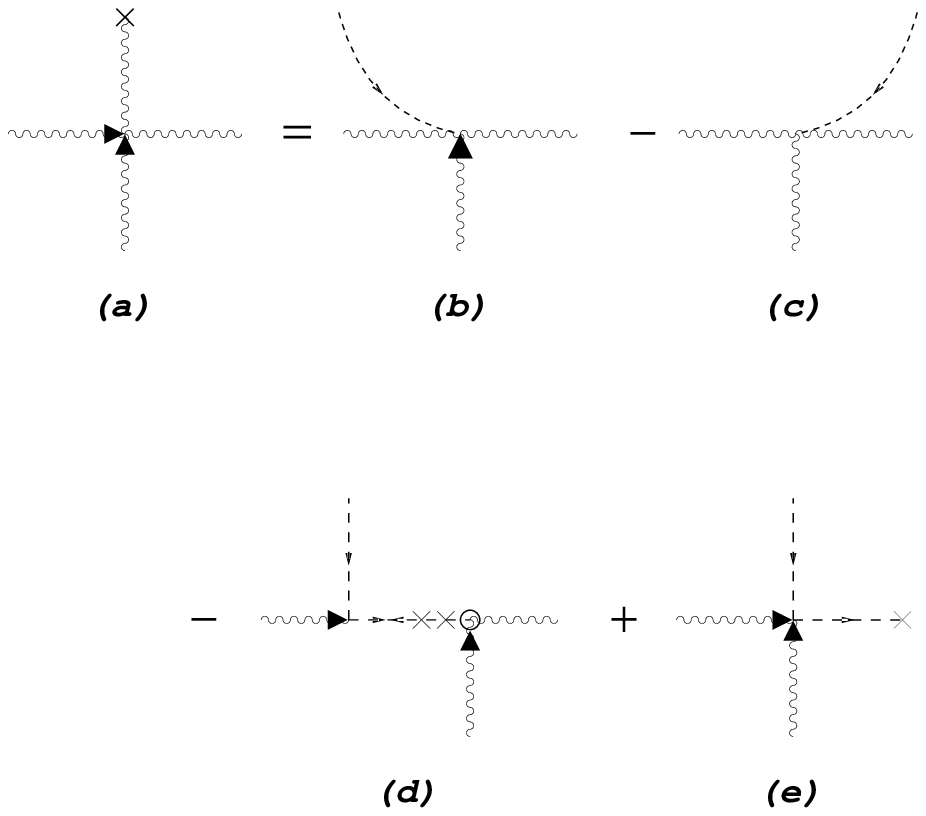}}
\nobreak
\vskip -9 cm\nobreak
\vskip .1 cm
\caption{Divergence relation of the four-gluon vertex 18c expressed in terms
of the BG vertices.}
\end{figure}

\begin{figure}
\vskip -0 cm
\centerline{\epsfxsize 4.7 truein \epsfbox {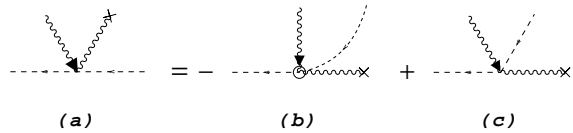}}
\nobreak
\vskip -13 cm\nobreak
\vskip .1 cm
\caption{Divergence relation of the ghost vertex 18e expressed in terms
of the BG vertices. The cross is adjacent to the arrowed line.}
\end{figure}

\begin{figure}
\vskip -3.5 cm
\centerline{\epsfxsize 4.7 truein \epsfbox {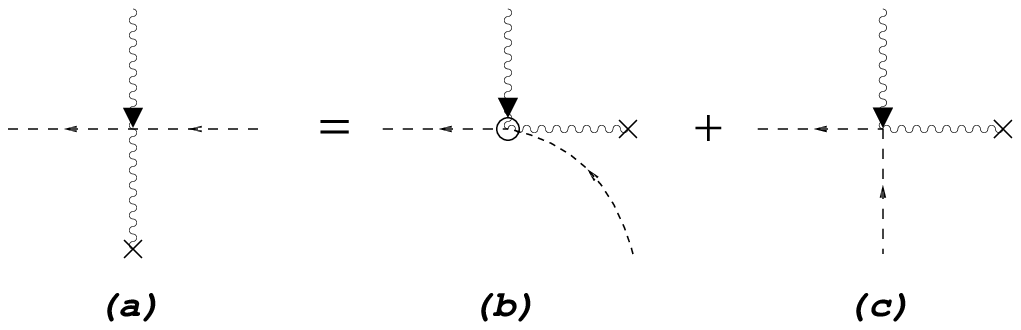}}
\nobreak
\vskip -10 cm\nobreak
\vskip .1 cm
\caption{Divergence relation of the ghost vertex 18e expressed in terms
of the BG vertices. The cross is opposite to the arrowed line.}
\end{figure}

The signs contained in this modified canonical structure are precisely
correct to make the extraneous diagrams cancel, extraneous meaning those
not needed to produce new BG vertices. For example, consider the combination
of diagrams 33a, 33b, and 33c. By using 3 and then 26, 33b produces 33e and
33f, and 33c produces 33g. Using 29, these combine to cancel, leaving behind
only the propagating diagram 33h to move on to other vertices.

\begin{figure}
\vskip -0 cm
\centerline{\epsfxsize 4.7 truein \epsfbox {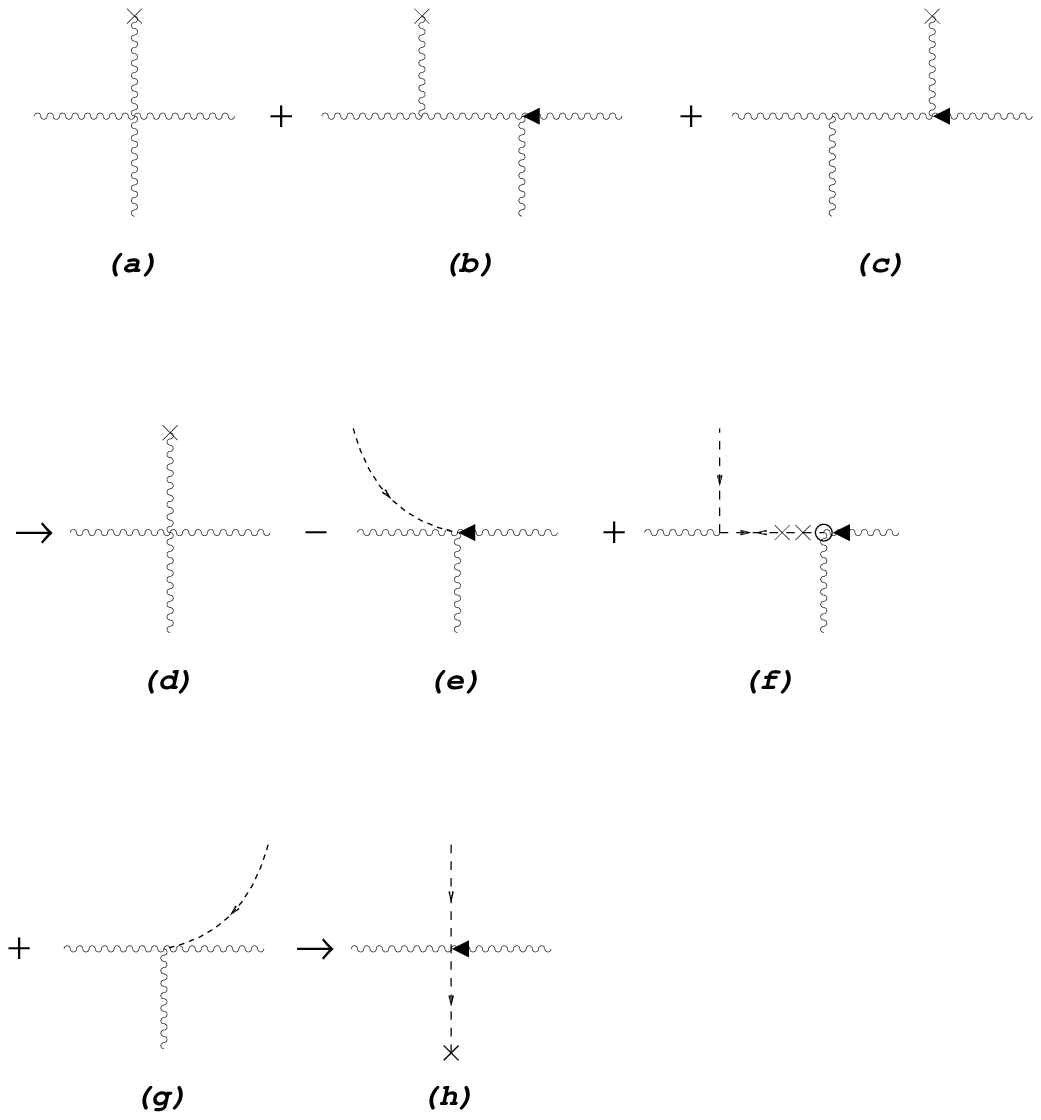}}
\nobreak
\vskip -8 cm\nobreak
\vskip .1 cm
\caption{An example of local gauge cancellations, leaving behind just
a propagating term to work on another vertex.}
\end{figure}

\begin{figure}
\vskip -0.8 cm
\centerline{\epsfxsize 4.7 truein \epsfbox {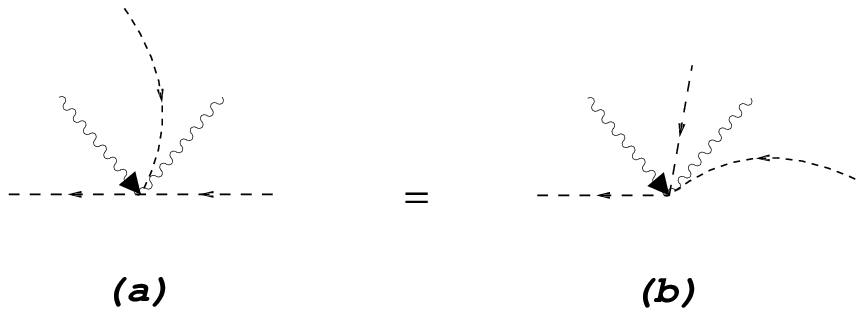}}
\nobreak
\vskip -14 cm\nobreak
\vskip .1 cm
\caption{Equality of a ghost line sliding into the ghost vertices 18e.}
\end{figure}

\begin{figure}
\vskip 0.5 cm
\centerline{\epsfxsize 4.7 truein \epsfbox {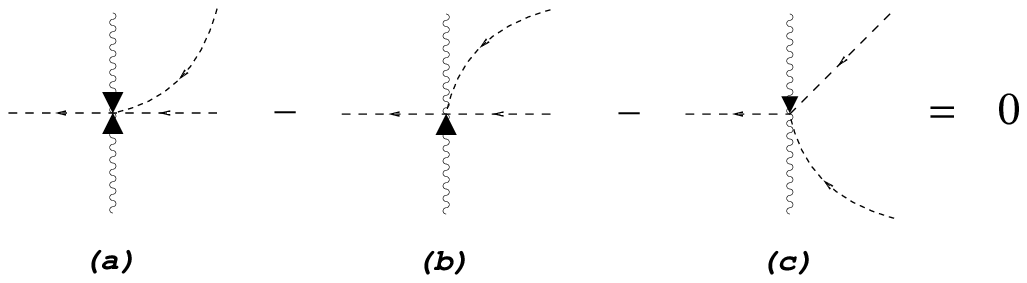}}
\nobreak
\vskip -14.8 cm\nobreak
\vskip .1 cm
\caption{Equality of a ghost line sliding into the ghost vertices.}
\end{figure}

\begin{figure}
\vskip -3.5 cm
\centerline{\epsfxsize 4.7 truein \epsfbox {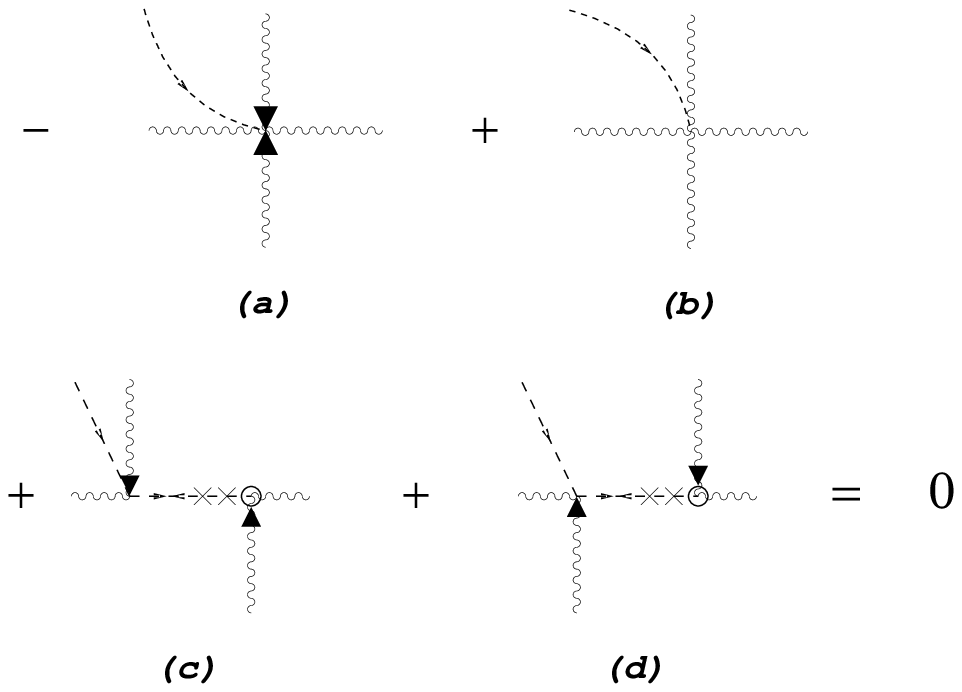}}
\nobreak
\vskip -8 cm\nobreak
\vskip .1 cm
\caption{Equality of a ghost line sliding into four-gluon vertices.}
\end{figure}

\begin{figure}
\vskip -0 cm
\centerline{\epsfxsize 4.7 truein \epsfbox {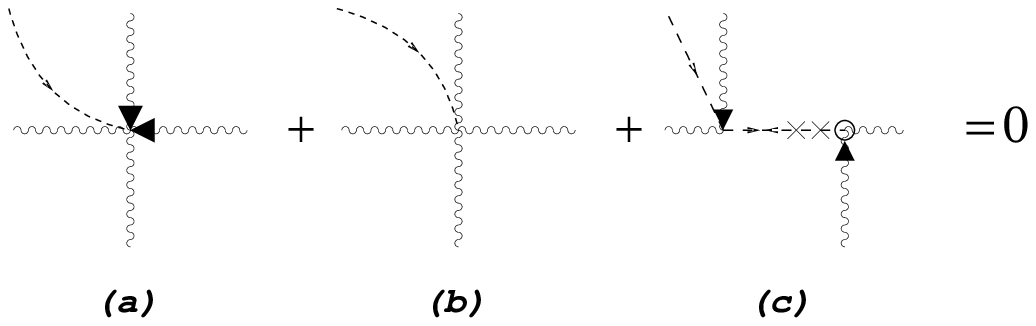}}
\nobreak
\vskip -13.5 cm\nobreak
\vskip .1 cm
\caption{Equality of a ghost line sliding into four-gluon vertices.}
\end{figure}

\begin{figure}
\vskip -0.5 cm
\centerline{\epsfxsize 4.7 truein \epsfbox {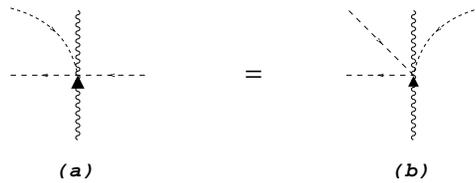}}
\nobreak
\vskip -13 cm\nobreak
\vskip .1 cm
\caption{Equality of a ghost line sliding into the ghost vertices 18e.}
\end{figure}

Similar cancellation takes place in all other cases, if we take into account
other identities shown in 34 to 38.

It is important to remark that in the
definition of the BG vertices 18 to 25, as well as in the divergence relations
26 to 32, all the external lines may be taken off-shell.

Let us now return to the construction of the BG vertices for $n=2$,
and consider the effect of the
cross 19b originating from the second vertex reaches the first (BG) vertex.
Because of the modified canonical structure of 26, things proceed
similarly to Lemma 1 and 2. Various situations can happen when this
cross returns to vertex 2 via a gluon loop. A sliding contribution just
before it returns to the second vertex gives rise to 21 to 23.

If the first vertex is just to the right of the second vertex, then in
using 19 to convert the second vertex, the contribution of 19c followed
by 26 on the first vertex produces terms like 24b, 25b, and 25c. When
combined with ordinary four-gluon vertex attached to these two external
lines, the BG vertices 24c and 25d are produced.

This completes the proof  of Lemma 3 for $n=2$. For $n>2$, it is easy
to see that nothing new can happen, in the sense that three or
more  vertices can interact only pairwise. This then completes the proof
of the Lemma for all $n$.

\section{Examples of new gauges}
The technique developed in \S II was used
in \S III to show the independence of the gauge used in
gluon propagators, and in
\S IV to convert ordinary to BG vertices. There is no reason why
it cannot be used to study other gauge problems, including our eventual
hope to find the `best' gauge for computing a specific set of
gauge-invariant diagrams.

In the present section, we will discuss
two unconventional gauges, one for four-gluon tree amplitude and the other for
two-loop gluon self-energy
diagrams. The new gauges carry fewer terms than either the ordinary
or the BG gauge. Though they may not be
the `best' gauge possible for the specific problems, nevertheless,
it does illustrate the fact
that improvements can be made on existing gauges using the techniques
developed in this paper.

The tree-diagram example is given in Fig.~39, where the arrowed vertices
are given in 18 (eq.~(4.1) and (4.4)). It can be shown using the techniques
above that a new
gauge containing the vertices in these diagrams does exist, {\it viz.,}
the three diagrams in 39 do sum up to give the four-gluon on-shell
scattering amplitude.

Note that this is {\it not} the BG gauge. Since all
the four lines in 39a are external, vertex 1d should
be used in the background gauge instead of 39a.
Moreover, each of the two 3g vertices in 39b,c
contains two external lines, again 1b rather than 18a should have been
used in the background gauge. In other words, there is no difference
between the BG and the ordinary gauge in this process.

Fig.~39a contains one term but the corresponding
vertex in BG contains 3 terms. Each of 39b,c contains $4^2=16$ terms
whereas the corresponding diagrams in BG each contains $6^2=36$ terms.
Thus the new gauge depicted in 39 saves a total of 42 terms out of
the 75 terms needed in the BG.

The two-loop example is given in Fig.~40, where for easy  drawing
solid lines are used to denote gluons. 40a and 40b are the two-gluon-loop
diagrams in BG, whereas 40c and 40d are the same diagrams in the new gauge.
Again the arrowed vertices are those displayed in 18. The two-loop
gluon self-energy contains a total of
18 diagrams in BG, but owing to incomplete
cancellation of the divergence terms when we shift gauges,
new vertices and new diagrams are generated
and a total of 26 diagrams appeared in this new gauge. Nevertheless,
the new gauge still contains few terms in total, because
the largest number of terms for each gauge already occurs in
the diagrams shown. The BG contains a total of 1242 terms, of which
1152 are contained in 40a,b. In contrast, the new gauge contains a
total of only 766 terms, of which 512 are contained in 40c and 40d.
In other words, 62\% more labour is required to compute the two-loop
self energy in BG than in the new gauge, and clearly even more labour
is required to compute it in the ordinary gauge.

Other details of this gauge and the two-loop computation will appear
separately.

\begin{figure}
\vskip -4.5 cm
\centerline{\epsfxsize 4.7 truein \epsfbox {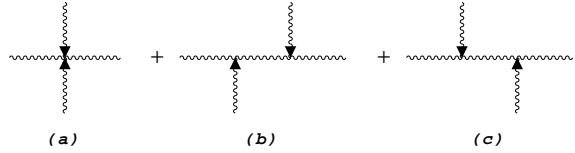}}
\nobreak
\vskip -9.5 cm\nobreak
\vskip .1 cm
\caption{Four-gluon tree amplitude in a new gauge.}
\end{figure}

\begin{figure}
\vskip -0 cm
\centerline{\epsfxsize 4.7 truein \epsfbox {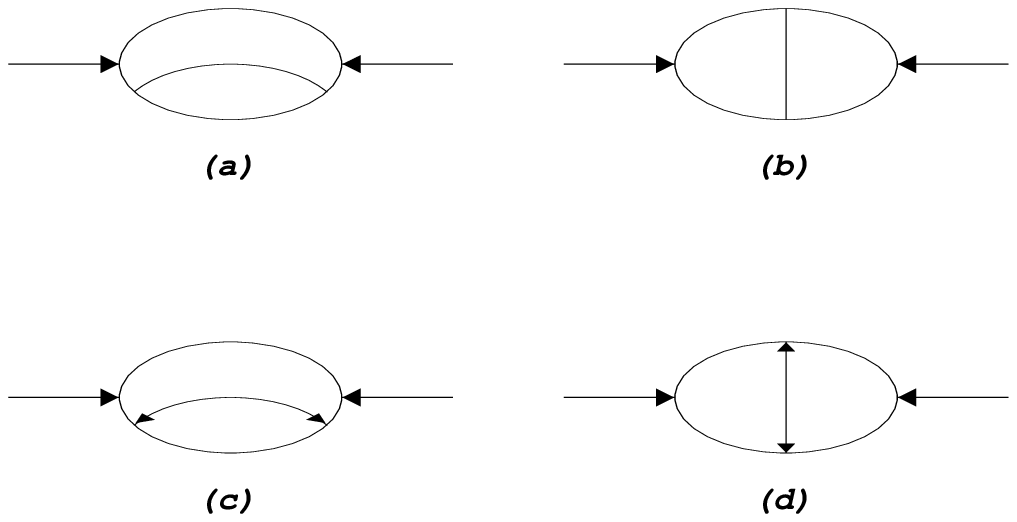}}
\nobreak
\vskip -12 cm\nobreak
\vskip .1 cm
\caption{Diagrams with two gluon loops for the gluon self-energy in the BG
gauge (a and b) and in the new gauge (c and d).}
\end{figure}

\section{Acknowledgments}
This research is supported in part by the Natural Science and Engineering
Research Council of Canada and by the Qu\'ebec Department of Education.
YJF acknowledges the support of the Carl Reinhardt Major
Fellowship. CSL thanks F. Wilczek for discussions
and hospitality at the Institute for Advanced Study in Princeton
where part of this manuscript was written.

\end{document}